\documentclass[11pt]{amsart}
\usepackage[english]{babel}
\usepackage[margin=1in]{geometry}

\usepackage{amssymb, amsmath, amscd, amsthm, color, epsfig, url, amsfonts, latexsym, graphicx, wrapfig, hyperref, multicol, wasysym, xcolor, framed, array, varwidth, caption, booktabs, tikz-cd}
\usepackage[color=blue!20, textwidth=1in, textsize=footnotesize]{todonotes}
\newcommand{\R}{{\mathbb R}}

\usepackage[margin=0.2cm]{caption}
\usepackage{subfig}
\usepackage{float}
\usepackage{multicol}
\usepackage{makecell}

\makeatletter
\renewcommand\l@subsection{\@tocline{2}{0pt}{2pc}{5pc}{}}
\makeatother

\begin{document}
	\title[TDA and MICS]{Topological data analysis and UNICEF Multiple Indicator Cluster Surveys}
	\author{Jun Ru Anderson}
	\address{Department of Mathematics, Wellesley College, 106 Central Street, Wellesley, MA 02481}
	\email{junru.anderson@wellesley.edu}
	\urladdr{andersonic.github.io}

	\author{Fahrudin Memi\'c}
	\address{UN Building, Zmaja od Bosne bb, 71000 Sarajevo, Bosnia-Herzegovina}
	\email{fmemic@gmail.com}
	\urladdr{linkedin.com/in/fahrudin-memi\'c-89927a106}
	
	\author{Ismar Voli\'c}
	\address{Department of Mathematics, Wellesley College, 106 Central Street, Wellesley, MA 02481}
	\email{ivolic@wellesley.edu}
	\urladdr{ivolic.wellesley.edu}
	
\keywords{Multiple indicator cluster surveys, MICS, wealth index, topological data analysis, Mapper algorithm, Mapper graph, UNICEF}


	\begin{abstract}
	
	Multiple Indicator Cluster Surveys (MICS), supported by UNICEF, are one of the most important global household survey programs that provide data on health and education of women and children. We analyze the Serbia 2014-15 MICS dataset using topological data analysis which treats the data cloud as a topological space and extracts information about its intrinsic geometric properties. In particular, our analysis uses the Mapper algorithm, a dimension-reduction and clustering method which produces a graph from the data cloud. The resulting Mapper graph provides insight into various relationships between household wealth -- as expressed by the wealth index, an important indicator extracted from the MICS data -- and other parameters such as urban/rural setting,  ownership of items, and prioritization of possessions. Among other uses, these findings can  serve to inform policy by providing a hierarchy of essential amenities. They can also potentially be used to refine the wealth index or deepen our understanding of what it captures. 	
	\end{abstract}
	

	\maketitle

	\tableofcontents
	
	\baselineskip=13pt
	\parskip=8pt
	\parindent=0cm
	

	\section{Introduction}



	

	
	Multiple Indicator Cluster Surveys, MICS (\url{mics.unicef.org}), are one of the most significant global sources of household data on health, education, and the well-being of women and children.  Supported by UNICEF, they have been conducted since the mid-1990s in over 100 countries. The data is gathered through face-to-face interviews in nationally representative samples of households and can be disaggregated in various ways. MICS occurs in multiyear rounds and provides tools and guidance for governments and institutions to create, inform, and implement socio-economic and health policies. MICS data and documentation are freely available at the MICS website. More details can be found in the survey article \cite{KH:MICS}.

	One important parameter that can be calculated from the MICS surveys and attributed to each household is its \emph{wealth index}. This number essentially captures the household wealth based on the ownership of certain items and is an important and widely used instrument for assessing the economic situation of a country. The calculation is performed through standard principal component analysis (PCA) of the data, with the wealth index depending on the first principal component. 	

	The goal of this paper is to analyze the MICS data and the wealth index with \emph{topological data analysis} (TDA). This is a relatively new technique that tries to extract intrinsic information from the shape of a data cloud and interpret this information as features of the data. For overviews of TDA and its applications, see~\cite{Carlsson:TDA, Carlsson:TDAHomotopy, OPTGH:TDAOverview}. 
  
  There are several successful TDA methods currently in use, and the one we will use in this paper is the \emph{Mapper algorithm}, due to Singh, M\'emoli, and Carlsson \cite{SMC:Mapper}. 	Mapper is an unsupervised machine learning algorithm which is essentially a dimension reduction and clustering procedure.
	The idea is to reduce the data to a \emph{Mapper graph} that retains various topological features of the data cloud. The nodes in the graph represent clusters of points that are ``nearby'' according to some notion of distance, while edges represent overlaps of clusters.  Because of its ability to retain the ``shape'' of data and hold both local (nodes) and global (edges) information, this algorithm seems to capture more information than other dimension-reduction procedures like PCA. At the same time, it is a powerful visualization tool since it reduces a high-dimensional cloud to a graph. 
	

	 Mapper has been used to great effect in a number of settings such as medicine, genomic analysis, neroscience, chemistry, remote sensing, soil science, agriculture, sports, voting, and economics (see \cite{BBBNL:Mapper} for a collection of  references). 
We believe, however, that the use of TDA to study standards of living and wealth inequality is novel.

	In more detail, we apply the Mapper algorithm to the 2014-15 MICS survey from Serbia.  The choice of this dataset is solely dependent on the fact that one of the authors (Memi\'c) was closely involved with creating, conducting, and analyzing the MICS surveys in Serbia and is thus intimately familiar with the methodology, in-field data collection, and the post-survey statistical analysis. The fact that this paper focuses on Serbia is in many way secondary; the takeaway is that the Mapper algorithm offers a different point of view on the MICS data and that it can be used to inform policy-making. Our analysis can be performed on any MICS dataset and, in fact, a future comparative study could reveal which socio-economic indicators are specific to particular countries or regions of the world and which are more universal.


The data that goes into our Mapper graph is based on the yes/no answers to 34 survey questions (a subset of all the questions MICS asks) about material possessions. This data is endowed with a distance function that captures the idea that households that are ``close'' are those that have a similar set of possessions, and that some possessions are less common than others. This leads to a definition of a probability-based semimetric which does not appear to have been used in literature before. 

Once the Mapper is generated, we extract two sets of observations from it. One is obtained by overlaying information onto the Mapper, i.e.~by coloring the nodes according to the wealth index or ownership of particular items. This analysis  provides insight into relationships between wealth and rural/urban lifestyle, ownership of items, and certain types of households. This is potentially useful information in terms of understanding how to most efficiently raise the standard of living; it suggests, for example, that for the majority of people, household amenities are more important than gadgets. This is helpful in light of the fact that there is a long tail at the low end of the distribution of wealth scores (Figure \ref{fig:wscore_dist}), which indicates that the dataset contains a number of households with wealth -- and, one might therefore suspect, standard of living -- several standard deviations below the average. Our analysis provides some insight into  what possessions might most efficiently raise the standard of living for these households.


The other way we study the Mapper is by looking at its graph-theoretic properties, namely paths and flares. Studying the paths provides insight into relative priorities households have in terms of ownership of items. One of the advantages of Mapper is that it picks up non-monotonic relationships between material possessions even without the field evidence pointing to the potential existence of such a relationship. Flares, on the other hand, appear to inform our understanding of household categorization in a way that is more subtle than one based simply on income and assets.

	Because the Mapper performs local-to-global information extraction, looking at the same graph with different overlay colors allows us to see both overall trends and deviations from those trends clearly, without noise obscuring either. Both the ability to analyze boolean data and the application of overlay colors allow the Mapper to discern patterns in how different possessions relate to wealth as quantified by the wealth index. This expands the usefulness of the Mapper, and TDA more generally, as a hypothesis-generating method.

	It should be noted that our analysis relies on an existing measurement of wealth, namely the wealth index, so we do not expect to redefine it with our methods. However, since Mapper is a more a subtle data reduction technique than PCA, the standard methodology for computing the wealth index, one of the takeaways from the work here is that perhaps Mapper could serve as a more nuanced way of understanding wealth scores.




	\subsection{Organization of the paper}
	

In writing this paper, we have attempted to keep the technical exposition of the Mapper algorithm and the underlying topology to a minimum.
However, the reader would be aided by familiarily with basic statistics, including a cursory understanding of principal component analysis (PCA), as well as elementary notions of linear algebra and topology such as maps and metrics.

The paper is organized as follows:
\begin{itemize}
\item In Section \ref{S:Background}, we provide background on the Mapper algorithm (Section \ref{S:Mapper}) and and MICS surveys (Section \ref{S:MICS}), including a brief review of the wealth score and the wealth index.	
\item 
	 Serbia MICS data and the parameters used to produce our Mapper graph are presented in Section \ref{S:Methods}. In particular, Section \ref{S:SerbiaMICS} lists the questions whose binary answers provide the input for the Mapper and discusses the distribution of the wealth score, which has a long tail at the long end. Sections \ref{S:Metrics}, \ref{S:Filter}, and \ref{S:Clustering} give details on the metric, filter, and clustering  -- the choices that have to be made in order for the Mapper to be generated. As mentioned above, one feature of independent interest in this setup is our definition of probablility-based (semi)metric that does not appear to have been used before. 
\item 
Section \ref{S:Results} gives two Mappers, one generated from a probability-based semimetric and one from the Euclidean metric. The latter simply provides evidence that the former is more informative and worth studying. 
\item We discuss and analyze the Mapper in Section \ref{S:Discussion}. We first overlay various information onto the Mapper in  Section \ref{S:Overlay}. As mentioned above, this leads to various observations about the relationship between the wealth index and the urban/rural split (Section \ref{S:Urban/rural}) as well as  some more nuanced characterizations of households according to ownership of items (Section \ref{item_types}).  We then study some of the graph-theoretic properties of the Mapper in Section \ref{S:Graph}. Following various paths connecting its nodes turns out to inform our understanding of the relative priorities that households have in ownership of items (Section \ref{S:Paths}). Flares, which, roughly speaking, are paths that separate from the main body of the Mapper and end in a node of degree one, on the other hand provide insight into what kind of items are considered more luxury and which are more essential.
\item We summarize our finding briefly in Section \ref{S:Conclusions}.
\item Section \ref{S:Future} is meant to convey that the analysis performed here is just the first step in applications of TDA to the MICS data and that this approach has much potential. Various future directions of investigation are laid out, and include expanding the set of questions used to generate the Mapper as well as those that are superimposed on it, modifying the metric and the filter function according to different statistically significant parameters, and the application of this method to MICS datasets from countries other than Serbia.

\end{itemize}


	\subsection{Acknowledgments}
	
	The third author was partially supported by the Simons Foundation.



	\section{Background}\label{S:Background}
	


	\subsection{Mapper Algorithm}\label{S:Mapper}
	

	The Mapper algorithm allows us to visualize high-dimensional datasets as a graph.\footnote{Or more generally as a \emph{simplicial complex}; we will not need this more general version here.} The graph retains many of the geometric properties of the original data cloud, such as connectedness and the presence of holes, but is easier to analyze.
	
Starting with a data set $X\subset \R^n$ consisting of vectors in some high-dimensional Euclidean space $\R^n$, the user first specifies a dimension-reducing function $h\colon X\to\R^d$ called the  \emph{filter} or \emph{projection}. The filter usually has some statistical meaning, and it varies depending on the context and the desired features of the data that are to be exposed. The target space is more manageable since $d$ is typically much smaller than $n$.  Our filter will simply provide the number of affirmative answers to a series of questions (see Section \ref{S:Filter}).

The image of the filter is then covered by $m$ overlapping hypercubes, i.e.~products of intervals, where $m$ is chosen by the user. The degree of the overlap can also be selected. Then, for each hypercube $i$ (where $1\leq i\leq m$), the datapoints in the preimage of hypercube $i$ are clustered.

For the clustering to be performed, a distance function on the data is needed. The standard Euclidean distance is often used, but other metrics are also employed depending on the context of the analysis.  In fact, one only needs a \emph{semimetric},  a distance function that does not necessarily satisfy the triangle inequality.  Our distance function will indeed be a semimetric  (see Section \ref{S:Metrics}). It will be based on probabilistic considerations and does not appear to have been used in the literature before.

With the notion of ``closeness'' in hand, one then uses some clustering algorithm, for example single-linkage hierarchical clustering, on the preimages of the filter map to decide which data points are close enough to be clustered together. Each cluster then becomes a node in the Mapper graph.
	
	Since the hypercubes overlap, clusters from different hypercubes may have datapoints in common. Two nodes have an edge between them precisely when this occurs. In other words, two nodes are connected by an edge if and only if the clusters they represent have non-empty intersection. 
	
\begin{figure}[h]
		\captionsetup{font=small}
		\centering
		\includegraphics[width=0.6\columnwidth]{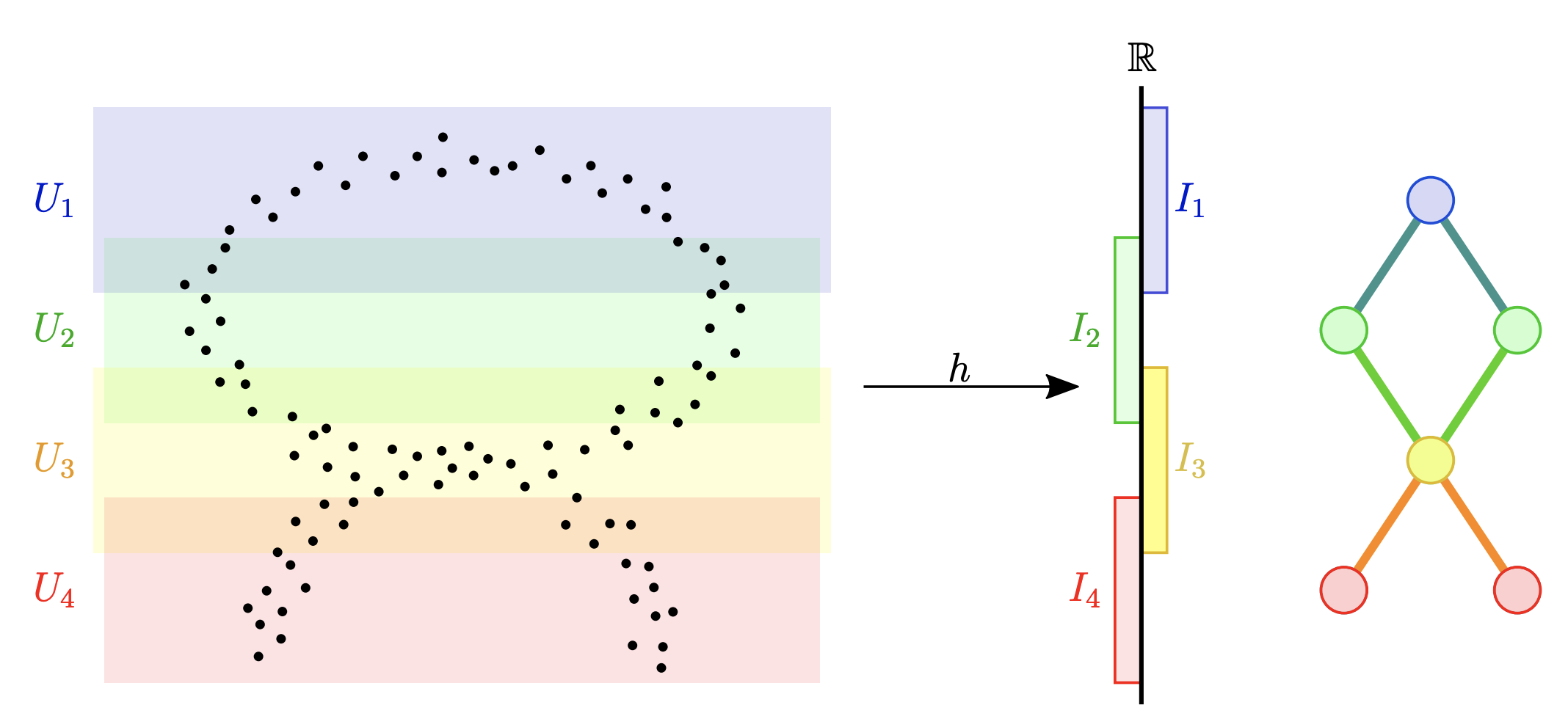}
		\caption{An illustration of the Mapper algorithm. The filter map $h\colon\R^2\to\R$ projects onto the $y$ coordinate. The image of $h$ is covered by 4 hypercubes (intervals in this case) $I_1,..., I_4$. The preimages of $h$ are the intersections of the open sets $U_1,..., U_4$ with the data cloud. These preimages are then clustered to form the nodes of the Mapper. An edge is created whenever a data point belongs to more than one cluster. Image source: Belch{\'i} et.al.~\cite{BBBNL:Mapper}.}
		\label{fig:MapperExample}
	\end{figure}

	Figure \ref{fig:MapperExample} illustrates the Mapper procedure on a simple dataset in $\R^2$, i.e.~each data point is in this case a vector with two components.
	
	The size of the nodes in the Mapper corresponds to how many data points are clustered in it.  The nodes can also be colored according to whatever attributes are deemed important. In our analysis, we will first color the node by the average wealth score of the households represented in the clusters, and then compare the result to the graph colored by rates of ownership of different possessions.
	
A number of free implementations of the Mapper algorithm exist, such as
Python Mapper~\cite{PythonMapper}, TDAmapper (R implementation)~\cite{TDAmapper}, and Kepler
Mapper \cite{KeplerMapper2019}. The last one is what we will use for our analysis.
	
Mapper is a data reduction method, and it is in this way similar to principal component analysis.  However, while the nodes are created from local information, the edges of the graph retain some knowledge about the global topological features of the data as a  subspace of $\R^n$.  This ability for local-to-global extrapolation is what makes Mapper appealing and useful.

It should be noted that the user makes a number of choices when implementing Mapper -- filter function, number of sets in the cover, the size of the overlap, metric, and clustering procedure -- and varying these parameters can produce very different graphs.

%
%
	

	\subsection{Multiple Indicator Cluster Surveys}\label{S:MICS}
	

	Since its inception in 1995, the Multiple Indicator Cluster Surveys (MICS) have become the largest source of data on women, children, and adolescents worldwide. Supported and conducted by UNICEF, the program is currently in its sixth multi-year round (MICS6). Over two decades, 341 MICS surveys have been carried out in 117 countries, helping shape policies for the improvement of the well-being of women and children. MICS was a major source of data on the UN Millennium Development Goals indicators and will continue to be a major data source for the UN 2030 Sustainable Development Agenda.
	
MICS surveys consist of a number of modules, some standard and some adapted to a specific country's needs.	Data is collected on a number of issues, such as fertility, mortality, contraception, newborn and mother health, and various other socio-economic parameters. Some of the modules are on the household level while others are on the individual level. 

The survey is conducted by trained fieldwork teams in face-to-face interviews with household members. The data is publicly available for research purposes from the MICS website.\footnote{See \url{https://mics.unicef.org}.}
	
An excellent overview of the MICS surveys, their methodology, history, and significance	can be found in \cite{KH:MICS}.\footnote{A number of papers discussing the MICS methodology can also be found at \url{https://mics.unicef.org/publications/reports-and-methodological-papers}.}


	\subsection{Wealth index and wealth score}\label{S:WI}
	
	

One important piece of information that is extracted from MICS surveys is the \emph{wealth index}.  This calculation captures the accumulated wealth by ranking households in terms of ownership of assets and amenities. 

A selection of MICS survey questions is used for the calculation of the wealth index. The questions are adapted to each country and are selected based on their perceived appropriateness for explaining the wealth of a household. In the case of Serbia that we consider in this paper, the subset of the 2014-15 MICS survey questions that were used for the computation of the wealth index had to do with material possessions and the features of the respondents' dwelling. 


The original wealth index calculation, used through MICS4, was modified in 2008 to take into account the urban bias. 
%
%
The current version involves the following steps:

\begin{enumerate}
\item	Select a set of variables that are thought be correlated with wealth. 
\item	Run separate principal component analyses (PCA) for urban and rural areas.
\item	Run a PCA for the whole population.
\item	Regress the urban and rural factor scores onto those for the general population.
\item	Obtain a combined \emph{wealth score}.
\item	Assign the combined wealth score to each household (give the same score to all household members).
\item	Divide the households into five equal groups, from poorest to richest, according to the combined wealth score, each containing 20 percent of the household members. The quintile in which a household ends up is its \emph{wealth index}.
\end{enumerate}

In a little more detail, principal components analysis (PCA), a data reduction technique, is at the heart of wealth index construction. From a set of variables that are correlated in terms of wealth, PCA extracts a set of uncorrelated principal components leading to the reduction of several variables in a data set into a smaller number of dimensions. 

Each dimension, or principal component, is a weighted linear combination of the initial variables that would best explain variance. In other words, each principal component is the sum of the variables multiplied by their weights. The weights for each variable are different in each principal component and are deduced from the data's correlation matrix.



The components are ordered so that the first principal component explains the largest amount of variation in the data. The wealth score calculation methodology uses the first principal component as the representation of wealth.




	\section{Methods}\label{S:Methods}
	


	\subsection{2014-15 Serbia MICS data}\label{S:SerbiaMICS}
	

	This paper focuses on the 2014-15 MICS5 data from Serbia, which participated in all six rounds of MICS (MICS6 data will be available by the end of 2020). The last three rounds were carried out on two independent samples in Serbia: nationally representative sample and the sample of the population living in Roma settlements. Analysis in this paper covers the data from Serbia MICS5 Survey on the nationally representative sample. 


Our analysis focuses on the following 34 questions: 
	\begin{multicols} {2}
		\begin{enumerate}
			\item Do you own an air conditioner?
			\item Do you own an animal-drawn cart?
			\item Do you have a bank account?
			\item Do you own a bed?
			\item Do you own a bicycle?
			\item Do you have cable tv?
			\item Do you own a car?
			\item Do you own a dishwasher?
			\item Do you own a drying machine?
			\item Do you own an electric stove?
			\item Do you have electricity?
			\item Do you own a freezer?
			\item Do you own a fridge?
			\item Do you own a hair dryer?
			\item Do you have internet?
			\item Do you own an iron?
			\item Do you own a microwave?
			\item Do you own a mobile phone?
			\item Do you own a motorcycle or scooter?
			\item Do you own a non-mobile phone?
			\item Do you own your dwelling?
			\item Do you own land that can be used for agriculture?
			\item Do you own animals?
			\item Do you own a pc/laptop?
			\item Do you own a radio?
			\item Do you own a table with chairs?
			\item Do you own a television?
			\item Do you own a tractor?
			\item Do you own a truck?
			\item Do you own a vacuum cleaner?
			\item Do you own a wardrobe?
			\item Do you own a washing machine?
			\item Do you own a watch?
			\item Do you own a water heater?
		\end{enumerate}	
	\end{multicols}
	
	These are all yes/no questions about material possessions. The answers to these questions are easy to encode and the TDA filter function is simple to define. One could in principle base the TDA analysis on different subsets of questions with different purposes in mind; we will make some comments about this in Section \ref{S:OtherData}.

	Of the $7351$ households surveyed, $6147$ households responded to all $34$ of the questions above, $1160$ households responded to none of these questions, and $44$ households responded to some, but  not all, of these questions. We confine our analysis to those households that answered all $34$ questions. The ``yes'' or ``no'' responses to each question are coded to values of $1$ and $0$, respectively. A household response to the questions will be a vector
	$$
	x= (x_1,...,x_{34})
	$$
whose coordinates are 0 or 1, with $x_i$ encoding the answer to the $i$th question. The dataset will thus consist of 6147 binary vectors of length 34.

	
	Much of our analysis will have to do with overlaying the wealth score data onto the Mapper graph resulting from the questions above.  The Serbia wealth score questions are broader than those, and, in addition to material possessions, they collect data on access to water, features of the dwelling, etc.
		
	The wealth score values of the  $6147$ households who responded to all of the 34 questions above range from $-7.68$ to $1.40$ with a mean of magnitude $<0.001$, a standard deviation of $0.99$ and a skew of $-2.48$. 
	
	\begin{figure}[h]
		\captionsetup{font=small}
		\centering
		\includegraphics[width=0.6\columnwidth]{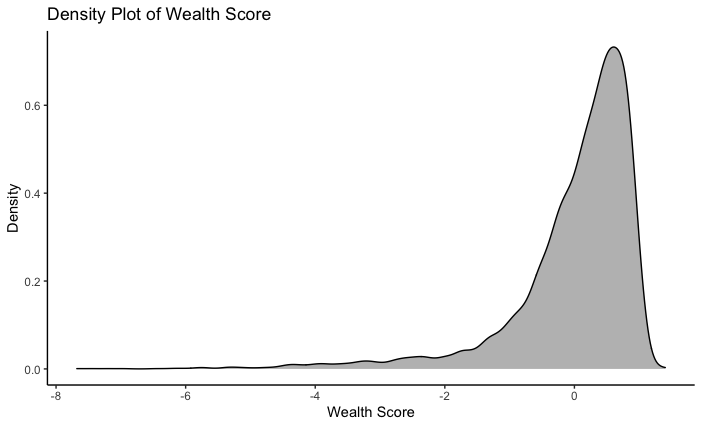}
		\caption{Serbia wealth score distribution}
		\label{fig:wscore_dist}
	\end{figure}
	
	\begin{figure}[h]
		\captionsetup{font=small}
		\centering
		\includegraphics[width=0.6\columnwidth]{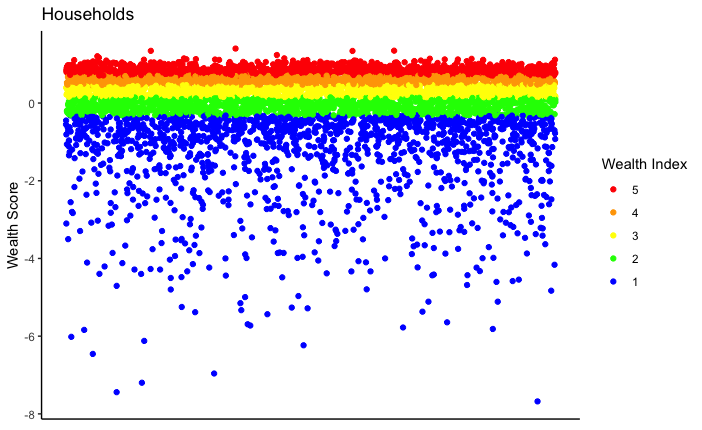}
		\caption{A color-coded visualization of Serbia wealth scores.}
		\label{fig:wscore_scatter}
	\end{figure}

As is evident from Figures \ref{fig:wscore_dist} and \ref{fig:wscore_scatter}, there is a long tail on the low end of the wealth score distribution; this can be seen in the skew of the distribution as well as by simply looking at the density plot. As mentioned in the Introduction, this is one of the features we believe TDA can give insight into.  In Section \ref{item_types}, we discuss how TDA can help identify which possessions (or lack thereof) distinguish the households in the tail. We also discuss how households can be classified with TDA in Section \ref{types_of_households}.
	
	
It should be noted that there is no corresponding tail on the high end of the distribution. It is possible that no such tail exists, but another explanation is that MICS survey questions do not differentiate between households that are moderately wealthy and households that are extremely wealthy. 


	
	\subsection{Metrics}\label{S:Metrics}
	

	As explained in Section \ref{S:Mapper}, in order to apply the Mapper algorithm to the Serbia MICS dataset $X$, we first need to define the distance $d(x,y)$ between two households' response vectors $x=(x_1,...,x_{34})$ and $y=(y_1,...,y_{34})$.

	We want the distance function to satisfy two intuitive properties:
	
	\begin{enumerate}
		\item If $x$ and $z$ agree wherever $x$ and $y$ agree, then $x$ and $z$ must be at least as close as $x$ and $y$. More formally, if 
		\begin{itemize}
		\item $A$ is the set of questions that households $x$ and $y$ agree on,
		\item $B$ is the set of questions that households $x$ and $z$ agree on, and 
		\item $A\subseteq B$, 		
		\end{itemize} 
	then
	$$
		d(x,z)\leq d(x,y).
		$$
		\item Uncommon similarities should be weighted more heavily. In other words, if many households own a television but fewer households own a car, two households both owning a car is a greater indicator of similarity than two households both owning a television.
	\end{enumerate}
	
	To satisfy these two conditions, we devise a new semimetric. We first compute, for each of the $34$ items, the proportion $p_i$ of households that answered yes to question $i$. This is hence the probability that a household answered yes to this question. Note that one could generate random response vectors from these probabilities by assuming independence among the different questions.
	
	We define the distance between two households to be zero if their responses are identical. Otherwise, we determine which questions the two households agree on and define the distance as the probability that two randomly generated response vectors would agree on these questions.
	
	In other words, if $x\neq y$, and $p=(p_1, ..., p_{34})$ is a vector of the proportions of households that own each of the $34$ items, we define
	\begin{align*}
	d\colon X\times X & \longrightarrow \R \\
	(x,y) & \longmapsto d(x,y) = \prod_{1\leq i\leq 34 \colon x_i=y_i}{p_i^2 + (1-p_i)^2}.
	\end{align*}
It is clear from the definitions that this satisfies the conditions of a semimetric, i.e.
\begin{itemize}
\item $d(x,y)\geq 0$,
\item $d(x,y)=0 \Longrightarrow x=y$,
\item $d(x,y)=d(y,x)$. 
\end{itemize}
However, this distance function is strictly a semimetric and not a metric, as it is possible that it violates the triangle inequality, i.e.~it is possible that $d(x,y)>d(x,z) +d(y,z)$ for some $x,y,z$. 

However, $d(x,y)$ does satisfy the two desirable conditions listed above:
	
	 Since $p_i$ is a probability, $p_i\in [0,1]$, and therefore $p_i^2 + (1-p_i)^2 \in [0,1]$. Then if $x$ and $y$ agree on questions $A$ and $x$ and $z$ agree on questions $B$ and $A\subset B$, we have
	
	\begin{align*}
	d(x,z) &= \prod_{i\in B}p_i^2+(1-p_i)^2 = \prod_{i\in A} p_i^2+(1-p_i)^2 \cdot \prod_{i\in \bar{A}\cap B} p_i^2+(1-p_i)^2 \\
	&=d(x,y) \cdot\prod_{i\in \bar{A}\cap B} p_i^2+(1-p_i)^2 \leq d(x,y) 
	\end{align*}
	
	Further, $p_i^2+(1-p_i)^2$, which is the probability that two random households from the dataset agree on question $i$, is a parabola with a minimum at $0.5$. Thus, a rarer similarity contributes a smaller factor to the product $d(x,y)$, which reduces the dissimilarity between two households more than a more common similarity.
	
Most of our analysis will use this probability-based semimetric (Section \ref{S:Probability}). However,	For comparison, we also also run the Mapper algorithm on the dataset $X$ with the standard Euclidean metric (Section \ref{S:Euclidean}).


	\subsection{Filter function}\label{S:Filter}

	For the filter function, we sum the components of each vector $x\in X$. This counts the number of items a household reported owning. Thus households that, at first glance, have a similar number of possessions are grouped together in the image. Then clusters within each open set in the image will be based on \textit{which} items a household owns rather than \textit{how many} items a household owns.
	
	Below is a table summarizing how many households reported owning any given number of things. For example,   this table says that 239 households reported owning 18 out of the 34 items surveyed or that 448 reported owing 27 items.
	\begin{center}
		\begin{tabular}{ |c|c|c|c|c|c|c|c|c|c|c|c|c|c|c|c|c|c|c|} 
			\hline
			$\mathbf{\sum_{i=1}^{34} x_i}$ & 1 & 2 & 3 & 4 & 5 & 6 & 7 & 8 & 9 & 10 & 11 & 12 & 13 & 14 & 15 & 16 & 17 & 18 \\
			\hline 
		
			{\bf Count} & 2 & 0 & 6 & 9 & 6 & 10 & 7 & 20 & 23 & 23 & 33 & 38 & 56 & 54 & 85 & 101 & 158 & 239 \\ 
			\hline
		\end{tabular}
	\end{center}

	\begin{center}
		\begin{tabular}{|c|c|c|c|c|c|c|c|c|c|c|c|c|c|c|c|c|}
			\hline
			$\mathbf{\sum_{i=1}^{34} x_i}$ & 19 & 20 & 21 & 22 & 23 & 24 & 25 & 26 & 27 & 28 & 29 & 30 & 31 & 32 & 33 & 34 \\
			\hline
			{\bf Count}  & 303 & 371 & 473 & 561 & 607 & 710 & 670 & 616 & 448 & 276 & 131 & 71 & 30 & 8 & 1 & 1 \\
			\hline
		\end{tabular}
	\end{center}

	To form the cover of the image of this filter function, we use $10$ intervals overlapping by $30\%$. Below is a table summarizing the cover of the image of the filter. The intervals are numbered 0-9. So for example, the first interval includes households that reported owning between one and five of the 34 items (inclusive), and there were 23 such households. It is important to keep in mind that the intervals overlap.
	
	\begin{center}
		\begin{tabular}{|c|c|c|c|c|c|c|c|c|c|c|}
			\hline
			{\bf Interval Number} & 0 & 1 & 2 & 3 & 4 & 5 & 6 & 7 & 8 & 9  \\
			\hline
			{\bf Interval Elements} & 1-5 & 4-8 & 7-11 & 11-14 & 14-18 & 17-21 & 21-24 & 24-28 & 27-31 & 30-34 \\
			\hline
			{\bf Household Count }& 23 & 52 & 106 & 181 & 637 & 1544 & 2351 & 2720 & 956 & 111 \\
			\hline
		\end{tabular}
	\end{center}


	\subsection{Clustering}\label{S:Clustering}
	

	We employ DBSCAN, a well-regarded general purpose clustering algorithm. DBSCAN requires two parameters: minPts and $\epsilon$ and classifies any datapoint $x$ with at least minPts points within distance $\epsilon$ (including $x$ itself) as a \emph{core point}. Any point $y$ within $\epsilon$ of $x$ is placed in the same cluster as $x$. If $y$ is itself a core point, any point within $\epsilon$ of $y$ is placed in the same cluster as $y$; this is applied recursively until all points are clustered, selecting a new $x$ whenever the edge of a cluster is reached. Any point that is not a core point and not in the cluster of a core point is classified as an outlier. \footnote{For more information on DBSCAN, see 
	\url{https://scikit-learn.org/stable/modules/generated/sklearn.cluster.dbscan.html}.
	}
	
\begin{figure}[h]
		\captionsetup{font=small}
		\centering
		\includegraphics[width=0.6\textwidth]{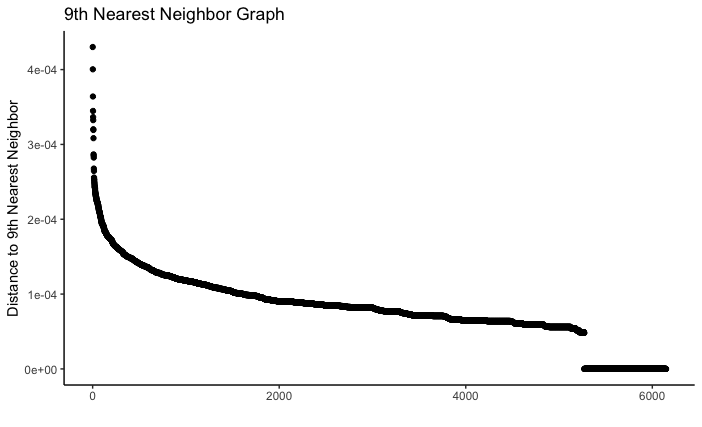}
		\caption{The parameter $\epsilon$ is chosen to be at approximately the bottom of the elbow.}
		\label{fig:epsilon}
	\end{figure}

	The choice of minPts is often guided by $\ln(n)$, where $n$ is the number of observations in the dataset. in this case, $n=6147$, the number of households who answered all 34 questions, and so $\ln(n) \approx 8.7$. We choose minPts to be $10$, so we wish to choose $\epsilon$ such that having $10$ points in a ball of radius $\epsilon$ suggests a cluster. To select $\epsilon$, we plot the distance from each point in the dataset to its $9^{\text{th}}$ nearest neighbor, as shown in Figure \ref{fig:epsilon}.
	
	This graph has a clear ``elbow.'' For distance values $d$ that are above the elbow, we expect that most points will have at least $9$ other points within $d$, yielding clusters that are too coarse. On the other hand, for values of $d$ that are below most of the graph, we expect very few points to have at least $10$ points within distance $d$; we would thus have very few core points and many points classified as outliers. We therefore choose $\epsilon$ to be at approximately the bottom of the elbow of the graph, i.e.~we choose $\epsilon = 10^{-4}$.
	
\begin{figure}[h]
		\captionsetup{font=small}
		\centering
		\includegraphics[width=0.6\textwidth]{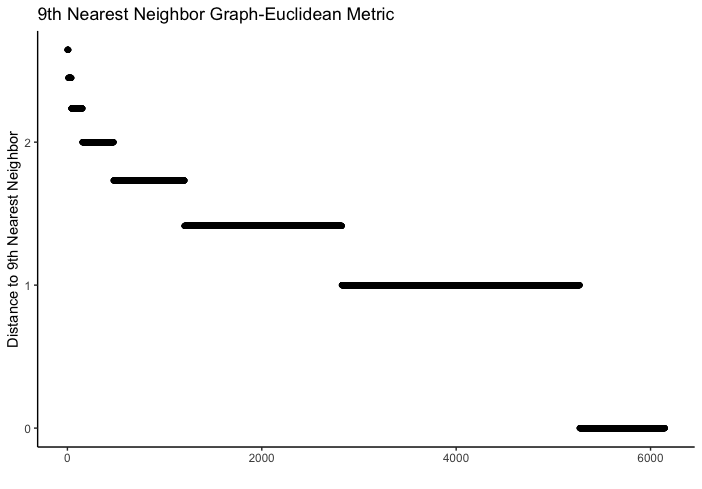}
		\caption{The parameter $\epsilon$ is normally chosen to be at the bottom of the ``elbow.''}
		\label{fig:EuclideanNNPlot}
	\end{figure}

	For the Euclidean metric, we can repeat this process. To select $\epsilon$, we once again plot the distance from each point in the dataset to its 9th nearest neighbor, as shown in Figure \ref{fig:EuclideanNNPlot}.
	
	It is harder to discern a clear elbow. Because data is binary, coordinates are either $0$ or $1$, and so the Euclidean metric produces only a discrete set of possible distances between points. We use $1.5$ as $\epsilon$.
	


	\section{Results}\label{S:Results}
	


	

Before we look at the Mapper graphs, we make some notational conventions and initial observations.


A node $n$ from a given Mapper graph will be denoted by 
$(i,c)$, where $i$ is the interval the households in $n$ fall into and $c$ is the cluster (within the interval $i$) that households from $n$ were placed in. Note that the labeling of $c$ is arbitrary, while the labels $i$ are in ascending order according to the number of items that a household reported possessing.
	
	For example, node (3,0) refers to the node containing (or referring to) households from interval 3 (possessing between 11 and 14 of the 34 items surveyed) that were clustered by DBSCAN into interval 3's cluster 0. Node (3,1) refers to the node containing households from interval 3 that were clustered into interval 3's cluster 1. Node (2,0) refers to the node containing households from interval 2 that were clustered by DBSCAN into interval 2's cluster 0.
	
	
	As explained in Section \ref{S:Mapper}, edges exist between nodes when the clusters represented by the nodes overlap. Because each interval overlaps at most one other interval on each side (as we are mapping to one-dimensional space), clusters can only overlap with other clusters that come from the intervals either directly above or below them (for example, clusters from interval $1$ can only overlap clusters from intervals $0$ or $2$). Therefore, a node can only have an edge between itself and a node representing a cluster from the interval above or below. Note that clusters within a given interval are, by definition, disjoint, and so a node cannot have an edge between itself and another node representing a cluster from the same interval. Thus, (3,0) and (3,1) are mutually exclusive by definition (that is, no household can be in both), but (2,0) is not necessarily mutually exclusive with either (3,0) or (3,1).
	
	

	\subsection{Probability-based semimetric}\label{S:Probability}
	
\begin{figure}[h]
		\captionsetup{font=small}
		\centering
		\includegraphics[width=1\textwidth]{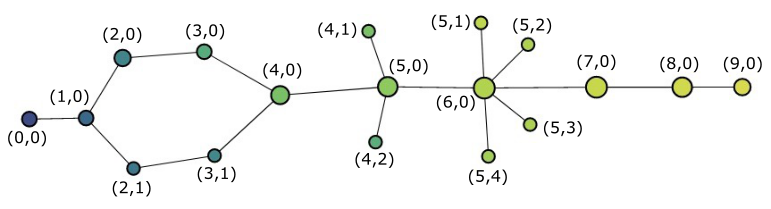}
		\caption{Mapper graph for the Serbia 2014-15 MICS data generated using the probability semimetric. Each node is colored based on the average wealth score of the constituent households. Darker colors indicate lower average wealth scores. The shape of the graph is irrelevant; the graph should be regarded up to isomorphism.}
		\label{fig:SerbiaMICS}
	\end{figure}

	Figure \ref{fig:SerbiaMICS} shows the Mapper graph generated using the semimetric described in Section \ref{S:Metrics} and the filter function from Section \ref{S:Filter}. Recall that each node represents a cluster of households that are contained in a single interval in the image.  Each node is colored based on the average wealth score of the households in the cluster that it represents, with purple corresponding to low average wealth score and yellow corresponding to high average wealth score. The size of the node reflects the number of households represented in it.

\begin{figure}[h]
		\captionsetup{font=small}
		\centering
		\includegraphics[width=0.8\textwidth]{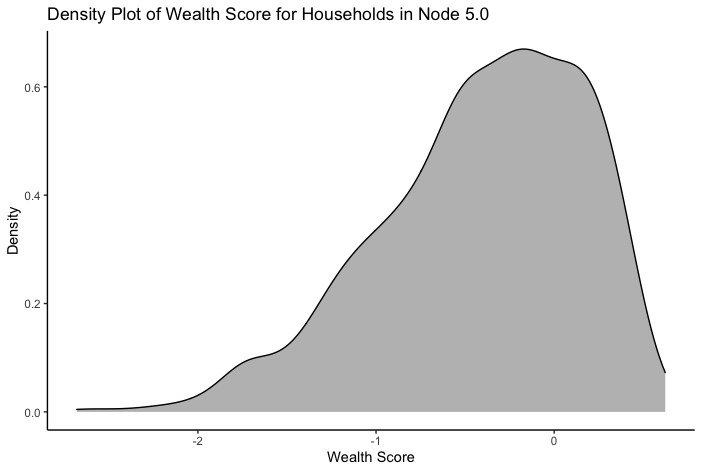}
		\caption{Distribution of the wealth score within node (5,0).}
		\label{fig:wscoredist}
	\end{figure}

	
As an example, Figure \ref{fig:wscoredist} gives the distribution of wealth scores for households in node (5,0).  	
	Table \ref{table:statspernode} summarizes key statistics of each node. 
	
	\begin{figure}[h]
		\begin{center}
			\begin{tabular}{|c|c|c|c|c|c|c|}
				\hline
				{\bf Node} & {\bf Average wscore} & {\bf wscore ${\mathbf \sigma}$} & {\bf wscore min} & {\bf wscore max} & {\bf Percent rural} & {\bf \# households}\\ \hline
				{\bf (0,0)}&$-5.747$&$0.937$&$-7.679$&$-4.145$&$0.652$&$23$\\ \hline
				{\bf (1,0)}&$-4.613$&$0.866$&$-7.441$&$-2.578$&$0.653 $&$49$\\ \hline
				{\bf (2,0)}&$-3.467$&$0.642$&$-4.795$&$-2.087$&$0.574$&$54$\\ \hline
				{\bf (2,1)}&$-4.002$&$0.588$&$-4.83$&$-2.983$&$0.929$&$14$\\ \hline
				{\bf (3,0)}&$-1.834$&$0.673$&$-3.29$&$-0.732$&$0.553 $&$38$\\ \hline
				{\bf (3,1)}&$-3.309$&$0.344$&$-3.984$&$-2.873$&$1.0$&$9$\\ \hline
				{\bf (4,0)}&$-0.842$&$0.665$&$-3.259$&$0.249$&$0.387$&$367$\\ \hline
				{\bf (4,1)}&$-0.836$&$0.43$&$-1.481$&$-0.194$&$0.333$&$9$\\ \hline
				{\bf (4,2)}&$-1.718$&$0.572$&$-3.08$&$-0.689$&$0.933$&$15$\\ \hline
				{\bf (5,0)}&$-0.405$&$0.573$&$-2.68$&$0.624$&$0.432 $&$1053$\\ \hline
				{\bf (5,1)}&$0.401$&$0.299$&$-0.348$&$0.586$&$0.0$&$9$\\ \hline
				{\bf (5,3)}&$0.105$&$0.388$&$-0.513$&$0.579$&$0.125$&$16$\\ \hline
				{\bf (5,2)}&$0.126$&$0.368$&$-0.576$&$0.53$&$0.067$&$15$\\ \hline
				{\bf (5,4)}&$-0.075$&$0.372$&$-0.778$&$0.301$&$0.333$&$9$\\ \hline
				{\bf (6,0)}&$0.2$&$0.451$&$-1.582$&$0.927$&$0.336 $&$1977$\\ \hline
				{\bf (7,0)}&$0.535$&$0.348$&$-0.988$&$1.141$&$0.349$&$2540$\\ \hline
				{\bf (8,0)}&$0.696$&$0.267$&$-0.207$&$1.236$&$0.468 $&$920$\\ \hline
				{\bf (9,0)}&$0.842$&$0.241$&$0.179$&$1.402$&$0.709 $&$103$\\ \hline
			\end{tabular}
		\end{center}
		\caption{Key statistics of each node. Here wscore denotes wealth score and $\sigma$ the standard deviation.}
		\label{table:statspernode}
	\end{figure}

	Note that the total number of households is $7220$. That is because some households appear in multiple nodes, due to the overlap of the intervals, while other households do not appear in any nodes because they were classified as outliers by DBSCAN.
	
	Tables \ref{table:itemspernode1} and \ref{table:itemspernode2} at the end of the article summarize how households in each node responded to each of the questions.
	
	
	\subsection{Euclidean metric}\label{S:Euclidean}
	
\begin{figure}[h]
		\captionsetup{font=small}
		\centering
		\includegraphics[width=0.6\textwidth]{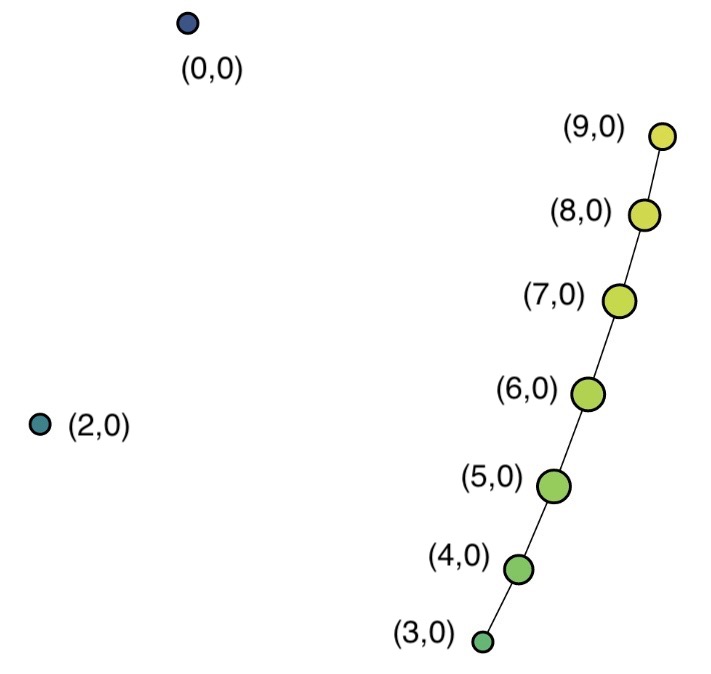}
		\caption{Mapper graph for the Serbia 2014-15 MICS data generated using the Euclidean metric. Each node is colored based on the average wealth score of the constituent households. Darker colors again indicate lower average wealth scores. The distance between the components and the shape of the graph are irrelevant}
		\label{fig:EuclideanMapper}
	\end{figure}

	Figure \ref{fig:EuclideanMapper} shows the Mapper graph generated using the Euclidean metric. As before, each node is colored based on the average wealth score of the household it represents. We see $9$ nodes, one for each interval except interval 1. This means that clustering using DBSCAN and the Euclidean metric only produced one cluster per interval. Note that this does not mean that every household in each interval was placed in a cluster. We can see from the lack of an edge between nodes (2,0) and (3,0) that the households in the intersection of intervals 2 and 3 were not in both of the respective clusters. We can also see from the lack of a node from interval 1 that no cluster was found in that interval.
	
	The fact that there is only one node for each interval means that the clustering diregards many of the interesting properties that we see in the graph based on the probability-based semimetric, such as the loop and flares off the trunk of the graph. The absence of these features makes the Euclidean graph less useful in terms of gaining insight into the relationship between possessions and the standard of living. In fact, one of the reasons we are considering the Euclidean metric-based graph is to illustrate the contrast with the probability-based semimetric and provide validation for using this more interesting distance function.
		
	 Another key feature to note is that while the graph in Figure \ref{fig:SerbiaMICS} is connected, the one in Figure \ref{fig:EuclideanMapper} is not. 
	

	\section{Discussion}\label{S:Discussion}
	

	\subsection{Overlaying information onto the Mapper}\label{S:Overlay}

	In this section, we will consider other information from the MICS survey and overlay them on the Mapper graph from Figure \ref{fig:SerbiaMICS} by coloring the nodes accordingly. This will lead to some insights about the relationship between the wealth score and living in urban or rural areas as well as the wealth score and ownership of certain types of items.

	\subsubsection{Wealth score and urban/rural living}\label{S:Urban/rural}

\begin{figure}[h]
		\captionsetup{font=small}
		\centering
		
		\subfloat[Rural households]{%
			\includegraphics[clip,width=0.6\columnwidth]{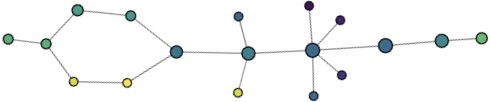}%
		}
		
		\subfloat[Land ownership]{%
			\includegraphics[clip,width=0.6\columnwidth]{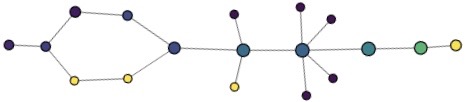}%
		}
		
		\subfloat[Tractor ownership]{%
			\includegraphics[clip,width=0.6\columnwidth]{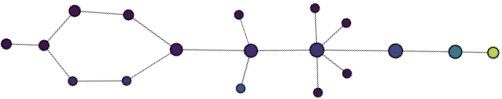}%
		}
		\caption{Graphs illustrating the relationship between wealth and rural lifestyle.}
		\label{fig:urbanrural}
	\end{figure}

	Figure \ref{fig:urbanrural} shows three figures where  the underlying graph is the Mapper from Figure \ref{fig:SerbiaMICS}, but now the nodes are colored according to the answers to particular questions.  In the graph (A), the nodes are colored based on the percentage of households that were classified as rural by the MICS study. Darker shades of purple indicate a lower proportion of urban households, while lighter shades of yellow indicate a higher proportion of urban households. The nodes in graphs (B) and (C) are colored based on land ownership and tractor ownership, two items closely related to rural farming. Darker shades of purple correspond to a lower percentage of households reporting owning the item in question, and lighter shades of yellow correspond to a higher percentage.

	It is common that people living in rural areas are, on average, less well-off than people living in urban areas. Figure \ref{fig:urbanrural}, for the most part, substantiates this belief. We see that the lowest-wealth nodes (referring to the colors in Figure \ref{fig:SerbiaMICS}) are among the lightest on the graph of rural households (Figure \ref{fig:urbanrural}(A)), and that nodes get progressively darker as we move towards the higher-wealth end of the graph. 
	
	This  is mirrored in the data on land ownership. While one might think that owning land is a sign of wealth, households that own land are significantly less wealthy than households that do not. (If we test this hypothesis, we get a $p$-value $<2.2\times 10^{-16}$. The $95\%$ confidence interval for the difference in wealth score between land-owning and non-land-owning households is $( -0.44, -0.38)$.) This can be explained, however, if living in a rural area mediates the relationship between land ownership and wealth (in other words, if owning land is correlated with living in a rural area and living in a rural area is correlated with lower wealth, that would explain the initially unintuitive fact that owning land is correlated with lower wealth). 
	
	There is, however, one notable exception: The last node on the right, which has the highest average wealth score, has (proportionately) more rural households than the several nodes to the left of it. These households are also much more likely to own land. One possibility is that, while city-dwellers are wealthier on average, that the wealthiest individuals are, in fact, in agriculture. The spike in tractor ownership in the last node supports this interpretation, since tractors are an agricultural tool. Another possible interpretation is that people, upon becoming wealthy, choose to move out of cities and into rural areas. This would not, however, fully explain why even the wealthiest urban people report owning land that can be used for agriculture, nor why many own tractors.
	
	\subsubsection{Wealth score and ownership of items}
	\label{item_types}
	
\begin{figure}[h]
		\captionsetup{font=small}
		\centering
		\begin{minipage}{.33\textwidth}
			\begin{figure}[H]
				\centering
				\subfloat[Electricity]{%
					\includegraphics[clip,width=0.9\columnwidth]{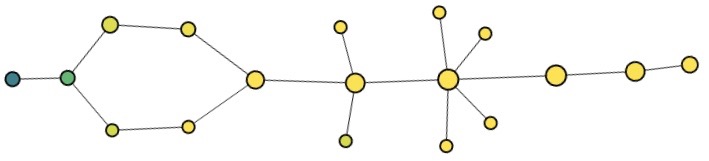}%
				}
				
				\subfloat[Bed]{%
					\includegraphics[clip,width=0.9\columnwidth]{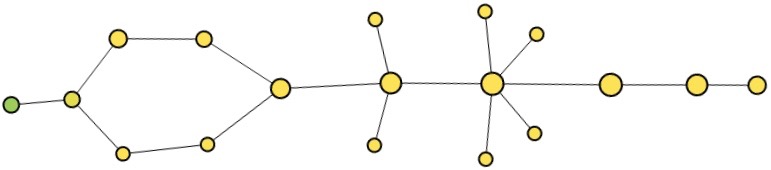}%
				}
				
				\subfloat[Table with chairs]{%
					\includegraphics[clip,width=0.9\columnwidth]{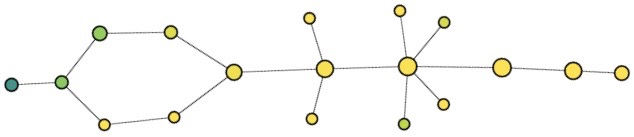}%
				}
				
				\caption*{Essentials}
			\end{figure}
		\end{minipage}%
		\begin{minipage}{.33\textwidth}
			\begin{figure}[H]
				\centering
				\subfloat[Car]{%
					\includegraphics[clip,width=0.9\columnwidth]{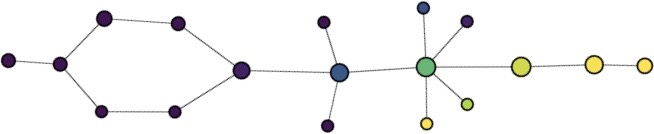}%
				}
				
				\subfloat[Cell phone]{%
					\includegraphics[clip,width=0.9\columnwidth]{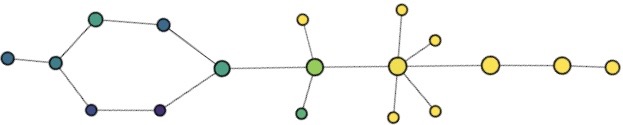}%
				}
				
				\subfloat[Computer]{%
					\includegraphics[clip,width=0.9\columnwidth]{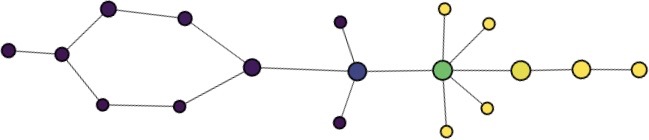}%
				}
				
				\caption*{$\longrightarrow$
				}
			\end{figure}
		\end{minipage}%
		\begin{minipage}{.33\textwidth}
			\begin{figure}[H]
				\centering
				\subfloat[Dishwasher]{%
					\includegraphics[clip,width=0.9\columnwidth]{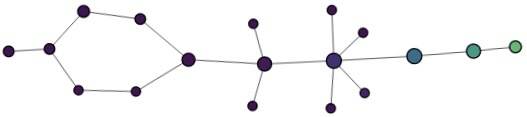}%
				}
				
				\subfloat[Motorcycle]{%
					\includegraphics[clip,width=0.9\columnwidth]{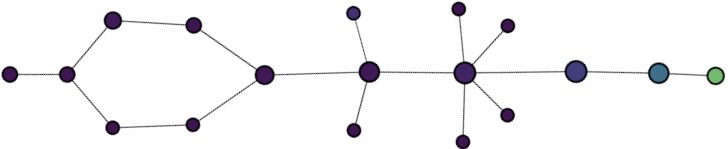}%
				}
				
				\subfloat[Drying machine]{%
					\includegraphics[clip,width=0.9\columnwidth]{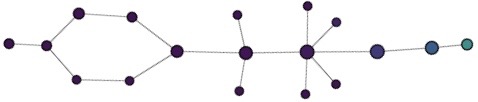}%
				}
				
				\caption*{Luxuries}
			\end{figure}
		\end{minipage}
		\caption{Mapper graphs of various possessions.  The arrow $\longrightarrow$ indicates a progression from items that are considered more essential to those that are considered more luxury.}
		\label{fig:possessions}
	\end{figure}

	We next investigate the relationship between certain possessions and overall wealth score by coloring the graph from Figure \ref{fig:SerbiaMICS} by the percentage of households that own a given item. Figure \ref{fig:possessions} shows nine graphs, divided into three categories. 
	As before, darker shades of purple correspond to a lower percentage of households reporting owning the item in question, and lighter shades of yellow correspond to a higher percentage.
	
	These graphs show three main classes of possessions. There are ``essential'' possessions, characterized by near-100\% owenership by all but the poorest of households.  There are ``luxury'' items, characterized by near-0\% ownership by all but the richest of households and relatively low rates of ownership by even those households.  And then there are the ``middle-class amenities,'' characterized by very low rates of ownership among the poorest households and a rather sudden jump to high rates of ownership around node (5,0) or (6,0).
	
	Note that this classification essential and luxury items is not the same as simply saying that the least-owned objects are luxury items. For example, animal-drawn carts are the single least-owned item, but this is not a luxury item. The ownership of this item is simply consistently low across nodes with varying average wealth scores.
	
\begin{figure}[h]
		\captionsetup{font=small}
		\centering
		
		\subfloat[Graph of TV Ownership]{%
			\includegraphics[clip,width=0.6\columnwidth]{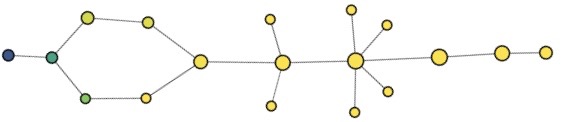}%
		}
		
		\subfloat[Graph of Microwave Ownership]{%
			\includegraphics[clip,width=0.6\columnwidth]{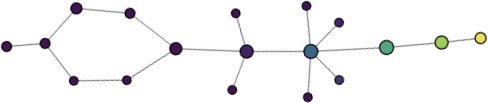}%
		}
		\caption{Mapper graphs of TV and microwave ownership}
		\label{fig:TVmicrowave}
	\end{figure}

	This classification method can be applied to all of the MICS questions about possessions, and reveals perhaps unintuitive categorizations for certain items. For example, Figure \ref{fig:TVmicrowave} shows two more colorings, this time by TV ownership and by microwave ownership.
	
	The graph of TV ownership most closely resembles that of an ``essential'' item, even though popular sentiment might not classify it as such (for example, people receiving government aid who purchase televisions might be criticized as being irresponsbile with their money). On the other hand, the graph for microwave ownership most closely resembles that of a ``luxury'' item, although many people in wealthy countries consider it essential. 
	
	The interpretation that people are spending their money unwisely is difficult to justify because a microwave is less expensive than a TV. In other words, a person or household that is able to purchase a TV is also able to purchase a microwave, so the fact that people tend to choose TVs over microwaves suggests that a television has a greater positive effect on happiness than a microwave. This also reminds us that cultural norms are not universally applicable; while someone in the United States might consider a microwave an essential tool for cooking, this appliance may not be as desirable in another country due to a difference in the style of food preparation.
	
	
	\subsubsection{Wealth score and types of households}
	\label{types_of_households}
	
	Nodes (0,0) and (1,0) are the poorest households. In these nodes, ownership of essential items is not universal. Though two-thirds of these households are classified as rural, many do not own their home and few own land.
	
	Nodes (2,1) and (3,1) exemplify the category of the "rural poor." These households all own land and most ($85.7\%$ in node (2,1) and $100\%$ in node (3,1)) own animals as well. They mostly own the essential items but very few, if any, middle-class amenities.
		
Middle-class and upper-class households can be classified based on item types. In other words, node (7,0) is the first node at which many luxury items such as air conditioners and microwaves appear in the majority of households, and so we can consider nodes (7,0), (8,0), and (9,0) to be comprised of upper-class households. Similarly, node (4,0) is where many of the middle-class amenities start to appear in the majority of households, and so we can classify nodes (4,0), (5,0), and (6,0) to be comprised of middle-class households. 
		
	It should be noted that designations such as ``middle-class,'' ``upper-class,'' or ``poor'' are imprecise and overlapping even with full information about a household's situation. Indeed, (3,1) and (4,0), as well as (6,0) and (7,0), have some overlap since there is an edge between them. 
	
		
	\subsection{Graph-theoretic properties of the probability-based Mapper}\label{S:Graph}
	
	
	In this section, we analyze some graph-theoretic properties of the Mapper from Figure \ref{fig:SerbiaMICS} such as existence of interesting paths and flares.
	
	
	\subsubsection{Paths}\label{S:Paths}
	
\begin{figure}[h]
		\captionsetup{font=small}
		\centering
		\includegraphics[width=0.8\textwidth]{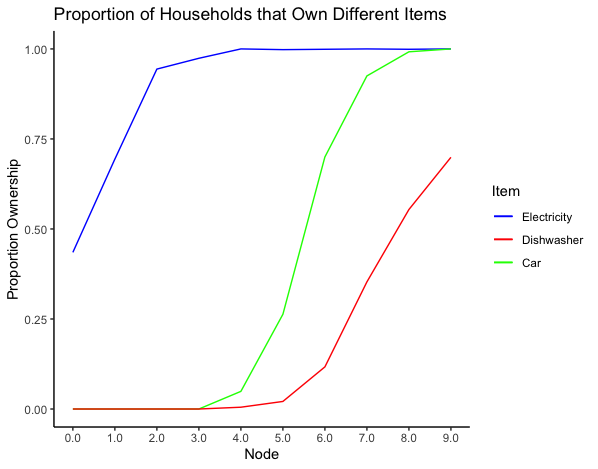}
		\caption{Graph of item ownership across different nodes in the path from node (0,0) to node (9,0) along nodes (2,0) and (3,0).}
		\label{fig:path1}
	\end{figure}

Item classification is reflected in the paths of Figure \ref{fig:SerbiaMICS}. This can be seen by  examining how item ownership changes as we move along the path. Figure \ref{fig:path1} shows the proportion of households in each node along the top path from node (0,0) to node (9,0) that own three different items/amenities: electricity (essential), a dishwasher (luxury), and a car (somewhere in between). This provides a different graphical representation of the information in the Mapper.
	
The word ``path'' here has a simple graph-theoretic meaning -- it is a sequence of edges that connect one node to another -- and  should not be understood as a path of economic mobility. While an argument could be made that some of these objects generate wealth (having a car might enable the commute to a higher-paying job), in most cases it is more likely that the item is a reflection of wealth (having a television requires a certain amount of wealth, but it does not  make its owner appreciably wealthier).
	
The merit of the analysis in Figure \ref{fig:path1} lies in providing a more precise picture of relative priorities. This graph shows that ownership of cars does not emerge until ownership of electricity is at nearly 100\%. On the other hand, while dishwasher ownership spikes later and a bit less steeply, we begin to see dishwasher ownership before car ownership reaches 100\%. From this, we can infer that people prefer electricity to cars and cars to dishwashers, but that the latter preference is not as strong or uniform.
	
	Analyzing the differences in the path that goes through nodes (2,0) and (3,0) and the path that goes through nodes (2,1) and (3,1) also highlights the urban/rural wealth difference. We see ownership of cell phones and televisions rise more quickly along the former path; this is discernible even from the Mapper graphs in Figure \ref{fig:possessions}. The difference in bed and electricity ownership is far less pronounced, and not immediately discernible from the Mapper graph's colors. This suggests that while the urban/rural wealth difference exists, in these households it manifests in items that are less essential than the life-critical ones. 
	


	\subsubsection{Flares}
	\label{flares}
	

We will not define precisely what we mean by a ``flare'' (one could do it in terms of node adjacencies and degrees) since this is clear from the graph in Figure \ref{fig:SerbiaMICS}:  Nodes (4,1) and (4,2) flare off from node (5,0), and nodes (5,1), (5,2), (5,3), and (5,4) flare off from node (6,0). It should be noted that these flares contain very few households --- node (5,3) contains the most, 16, and nodes (4,1), (5,1), and (5,4) contain the least, 9 each. Recall that nodes are generated by clustering households within each interval of the image of the filter function. As these flares are quite small relative to the number of households in the nodes that form the trunk of the graph, we can view them as households that are uncommon among those who answered in the affirmative to a similar number of questions.
	
	Node (4,2) looks very rural, with 100\% rates of land ownership and animal ownership. At first, it looks as though it might fall into the category of rural poor. Average wealth score is $-1.718$, compared to $-0.842$ for node (4,0) and $-0.837$ for node (4,1), continuing the trend of rural households having lower wealth than urban households in the same interval. 
	
	However, the average wealth score in node (4,2) is about 1.5 standard deviations higher than the average wealths core in node (3,1); a wealth score of $-3.309$ is in the tail end of the wealth score distribution, whereas a wealth score of $-1.718$ is on the border. Unlike the households in nodes (2,1) and (3,1), households in node (4,2) have a slew of amenities (such as electric stoves, irons, radios, vacuum cleaners, and washing machines) that are much less common among households in nodes (2,1) and (3,1). Further, five of the nine households in node (4,2) have a bank account, compared to none of the households in nodes (2,1) or (3,1). Thus, the (admittedly very few) households in node (4,2) might serve as a target for rural standard of living.
	
	
	Node (4,1), on the other hand, has 0\% land and animal ownership and an average wealth score similar to that of node (4,0). Instead, what sets node (4,1) apart from (4,0) and (4,2) is 100\% bike ownership (compared to $12.8\%$ for node (4,0), $0\%$ for node (4,2), and $33.7\%$ for node (5,0)), 100\% watch ownership (compared to $37.6\%$ in node (4,0), $0\%$ in node (4,2), and $57.6\%$ in node (5,0)), higher rates of mobile phone ownership, lower rates of non-mobile phone ownership and drastically lower rates of freezer ownership. We also see cable TV ownership at about half the rate of node (4,0) and one-third the rate of node (5,0). In broad terms, this looks like prioritization of items that are brought on one's person when going out every day over amenities that stay at home. This could in turn suggest a group of people who are prioritizing keeping up appearances. However, making this precise is difficult due to the small sample size. What we can infer, however, is that the overwhelming majority of households in interval 4 clustered into node (4,0), not node (4,1), and so the most people's preference for purchasing does prioritize household amenities. This is  useful information in terms of understanding how to most efficiently raise the standard of living -- it suggests that for the majority of people, household amenities are more important than the gadgets that are carried on one's person.

	Nodes (5,1)-(5,4) all have higher average wealth score than node (5,0) and are much less rural. All demonstrate 100\% laptop/computer ownership, compared to $18.8\%$ for node (5,0) and $73.8\%$ for node (6,0). Nodes (5,1), (5,2), and (5,3) also have much higher rates of TV ownership than node (5,0). Node (5,1) in particular is characterized by the average wealth score which is higher than in nodes (5,0), (5,2), (5,3), (5,4), and (6,0). We see no radios or bicycles and only one watch from the nine households, but universal owernship of ammenities such as air condinitioning, a luxury item, as well as cable TV and internet. As with node (4,1), the small number of households in node (5,1) make inference difficult. We can, however, say that the fact that these households clustered separately from the other households in node (5,0) confirms the luxury status of these amenities, since it indicates that households in interval 5 that have these amenities are ``far out'' from the main cluster of that interval.


	\section{Conclusions}\label{S:Conclusions}

	At the most general level, we have demonstrated that TDA can be used to investigate the shape of boolean data. In Euclidean space, the distance between two boolean vectors of length 34 can take on only 35 different values, which poses a challenge when attempting to investigate shape. By applying the probability-based semimetric, we allow the distances between two points to take on a much wider range of values while retaining the meaning of distance; this opens the door for clustering and TDA analyses to be performed.
	
	We have further demonstrated that using different overlay colors on the same Mapper graph can illuminate relationships between different variables. Because the Mapper performs local-to-global information extraction, looking at the same graph with different overlay colors allows us to see both overall trends and deviations from the overall trend clearly, without noise obscuring either. Both the ability to analyze boolean data and the application of overlay colors expand the usefulness of TDA as a hypothesis-generating method.

More specifically to the MICS data at hand, overlaying items on top of the Mapper graph and categorizing them accordingly can help guide which items or amenities ought to be prioritized for a possible  program that would supply them to individuals. Our ``essential'' items are defined by the fact that households with limited means chose to purchase them over middle-class amenities and luxury items. This pattern of purchasing would suggest that these items give higher marginal utility per unit cost, and that, given limited resources, providing these essential items to households might increase the utility of the population as a whole  in an efficient way.
	
	
	Household categorization could similarly be used to determine eligibility for certain types of aid. Governments or organizations could use our Mapper method to categorize households in a more nuanced way than simply stratifying based on income or assets. Such categorizations could help direct specific aid (for example, providing households with electricity or a table) to the households that would most benefit from it. This is valuable because more highly targeted aid may be cheaper (as it is provided to fewer households) and more politically feasible.


	\section{Future work}\label{S:Future}
	

	\subsection{Other MICS survey data}\label{S:OtherData} 	The MICS study collects much more data than the 34 questions we use for our analysis.  There are other questions relating to standard of living that do no elicit binary answers (for example, questions on roof composition and primary water source). Beyond this, there are questions on education level, vaccinations, access to health care, child labor, and more. These questions could be incoporated into the Mapper graph or overlayed on it as was done in Section \ref{S:Overlay}. This could, for example, reveal which households are more or less likely to vaccinate their children and guide the development of targeted policy to increase vaccination rates. 
	
	
However, in order to apply our probability-based metric to data with more than two answers, we must expand the notion of what it means for two households to agree on a given item. One straightforward approach that could work well with unordered categorical data would be to say that two households agree if they gave the same answer, and do not agree if they gave different answers. One could pre-compute the frequency of each response and then use either the probability that two households agree on the given item  (i.e.~$p_i^2$), or the probability that two households both gave the response (i.e.~$p_1^2 + p_2^2 + \cdots + p_n^2$).
	
	This approach would not work, however, for continuous data, and may be inappropriate for discrete ordered data. In this case, it might be better to create a distribution of the responses to a given question and use the probability that two randomly chosen households gave responses as close or closer than the two households in question.
	
	We could also use an \emph{isolation forest} to construct a lens for our data. Isolation forests quantify how ``unusual'' a datapoint is. If, for example, the anomalous households were primarily high wealth, that would suggest that there is a high tail in the distribution of wealth that the wealth score might not be capturing. If non-anomalous households form multiple clusters, this might indicate meaningful gaps between the lower, middle, and upper class. If they form a single cluster, that would suggest that there are households along the entire spectrum of socioeconomic status.
	
	
	
	\subsection{Other countries}
	

	As mentioned in the Introduction, we chose to work with the Serbia data because one of the authors participated in running  that MICS survey. However, UNICEF conducts MICS surveys in many different countries. Performing our analysis on data from different countries would allow one to draw region-specific insights, studying how people in different countries or regions prioritize their spending, and evaluating the validity and importance of the wealth score in policy-making decisions. It would also allow for generalizations of some insights to the global population if the same patterns emerged in various places around the world.

	
	\subsection{Other TDA methods}
	
	
	The Mapper algorithm is one of the two main topological data analysis tools. The other is \emph{persistent homology} \cite{Carlsson:TDA, Carlsson:TDAHomotopy} which regards the data cloud as a topological space and then tries to understand its topologically important features, namely those that are unchanged by continuous deformations. These include connectedness and existence of holes or ``voids'' of various dimensions. 
	
	The data cloud is first endowed with a metric (Euclidean, correlation, or anything else) that turns the data cloud into a space, and then one applies \emph{homology}, a standard algebraic tool in topology, to study it. The homology is calculated at different scales, and the features that ``persist'' at various levels are considered topologically meaningful and provide insight into the data cloud.
	
	Persistent homology has been used in a variety of settings (for an overview, see \cite{OPTGH:TDAOverview}) and is applicable to the setting of the MICS data. If this analysis for example exhibits the data cloud as a number of disconnected components, this might reflect disparities in wealth according to conditions or paramaters that are otherwise not easily discernible by the usual statistical methods.  If the data shows a presence of holes, this might indicate combinations of possessions that are not represented in the data. Identifying why any such combination did not appear (are the items redundant, is one useful only in a rural setting and another only in an urban one, etc.) could illuminate something about the landscape of consumer demand.
	

	\newpage 
	
		\begin{figure}[p]
		\begin{center}
				\begin{tabular}{|c|c|c|c|c|c|c|c|c|c|}
					\hline
					& {\bf (0,0)} & {\bf (1,0)} & {\bf (2,0)} & {\bf (2,1)} & {\bf (3,0)} & {\bf (3,1)} & {\bf (4,0)} & {\bf (4,1)} & {\bf (4,2)}\\
					\hline 
					{\bf Air conditioner}&0.0&0.0&0.0&0.0&0.0&0.0&0.022&0.0&0.0\\ \hline
					{\bf Animal-drawn cart}&0.043&0.02&0.0&0.0&0.0&0.0&0.008&0.0&0.0\\ \hline
					{\bf Bank account}&0.13&0.102&0.204&0.0&0.237&0.0&0.504&0.889&0.533\\ \hline
					{\bf Bed}&0.826&0.959&1.0&1.0&1.0&1.0&1.0&1.0&1.0\\ \hline
					{\bf Bicycle}&0.087&0.143&0.056&0.0&0.0&0.0&0.128&1.0&0.0\\ \hline
					{\bf Cable TV}&0.0&0.0&0.0&0.0&0.0&0.0&0.229&0.111&0.0\\ \hline
					{\bf Car}&0.0&0.0&0.0&0.0&0.0&0.0&0.049&0.0&0.0\\ \hline
					{\bf Dishwasher}&0.0&0.0&0.0&0.0&0.0&0.0&0.005&0.0&0.0\\ \hline
					{\bf Drying machine}&0.0&0.0&0.0&0.0&0.0&0.0&0.003&0.0&0.0\\ \hline
					{\bf Electric stove}&0.0&0.163&0.463&0.143&0.816&0.333&0.954&0.889&0.867\\ \hline
					{\bf Electricity}&0.435&0.694&0.944&0.929&0.974&1.0&1.0&1.0&0.933\\ \hline
					{\bf Freezer}&0.0&0.02&0.222&0.5&0.658&0.778&0.708&0.111&1.0\\ \hline
					{\bf Fridge}&0.087&0.347&0.685&0.643&0.974&1.0&0.995&1.0&1.0\\ \hline
					{\bf Hair dryer}&0.0&0.041&0.13&0.0&0.263&0.0&0.76&0.556&0.2\\ \hline
					{\bf Internet}&0.0&0.0&0.0&0.0&0.0&0.0&0.0&0.0&0.0\\ \hline
					{\bf Iron}&0.0&0.102&0.259&0.0&0.658&0.222&0.94&0.889&0.6\\ \hline
					{\bf Microwave}&0.0&0.0&0.0&0.0&0.0&0.0&0.0&0.0&0.0\\ \hline
					{\bf Mobile phone}&0.348&0.429&0.574&0.214&0.342&0.111&0.597&1.0&0.667\\ \hline
					{\bf Motorcycle or scooter}&0.0&0.0&0.019&0.0&0.0&0.0&0.014&0.111&0.0\\ \hline
					{\bf Non-mobile phone}&0.0&0.082&0.204&0.5&0.711&0.778&0.834&0.333&0.733\\ \hline
					{\bf Owns dwelling}&0.391&0.531&0.667&0.929&0.895&0.889&0.828&0.889&0.933\\ \hline
					{\bf Owns land}&0.087&0.163&0.037&1.0&0.211&1.0&0.218&0.0&1.0\\ \hline
					{\bf Owns animals}&0.0&0.041&0.056&0.857&0.079&1.0&0.144&0.0&1.0\\ \hline
					{\bf PC/laptop}&0.0&0.0&0.019&0.0&0.026&0.0&0.025&0.0&0.0\\ \hline
					{\bf Radio}&0.043&0.184&0.259&0.143&0.105&0.222&0.643&0.778&1.0\\ \hline
					{\bf Table with chairs}&0.522&0.796&0.815&1.0&0.947&1.0&0.978&1.0&1.0\\ \hline
					{\bf Television}&0.261&0.592&0.926&0.786&0.947&1.0&0.986&1.0&1.0\\ \hline
					{\bf Tractor}&0.0&0.0&0.0&0.071&0.0&0.111&0.038&0.0&0.267\\ \hline
					{\bf Truck}&0.0&0.0&0.019&0.0&0.0&0.0&0.005&0.0&0.0\\ \hline
					{\bf Vacuum cleaner}&0.0&0.0&0.074&0.0&0.421&0.111&0.905&0.889&0.733\\ \hline
					{\bf Wardrobe}&0.435&0.735&0.87&0.857&0.974&0.889&0.995&1.0&1.0\\ \hline
					{\bf Washing machine}&0.0&0.02&0.148&0.071&0.789&0.111&0.847&1.0&0.667\\ \hline
					{\bf Watch}&0.043&0.163&0.037&0.143&0.026&0.111&0.376&1.0&0.0\\ \hline
					{\bf Water heater}&0.0&0.02&0.333&0.143&0.842&0.222&0.918&1.0&0.8\\ \hline
				\end{tabular}
		\end{center}
		\caption{Proportion of households in nodes (0,0)-(4,2) answering each of the 34 questions in the affirmative.}
		\label{table:itemspernode1}
	\end{figure}
			
			\newpage
			
			\begin{figure}[p]
			\begin{center}
				\begin{tabular}{|c|c|c|c|c|c|c|c|c|c|}
					\hline
					& {\bf (5,0)} & {\bf (5,1)} & {\bf (5,2)} & {\bf (5,3)} & {\bf (5,4)} & {\bf (6,0)} & {\bf (7,0)} &  {\bf (8,0)} & {\bf (9,0)}\\
					\hline
					{\bf Air conditioner}&0.052&1.0&0.067&0.125&0.0&0.31&0.565&0.71&0.816\\ \hline
					{\bf Animal-drawn cart}&0.005&0.0&0.0&0.0&0.0&0.012&0.014&0.026&0.068\\ \hline
					{\bf Bank account}&0.749&0.889&0.733&1.0&0.556&0.907&0.964&0.978&0.981\\ \hline
					{\bf Bed}&1.0&1.0&1.0&1.0&1.0&1.0&1.0&1.0&1.0\\ \hline
					{\bf Bicycle}&0.337&0.0&0.067&0.062&0.222&0.526&0.782&0.905&0.951\\ \hline
					{\bf Cable TV}&0.375&1.0&0.867&1.0&0.0&0.701&0.829&0.893&0.893\\ \hline
					{\bf Car}&0.263&0.222&0.067&0.875&1.0&0.7&0.925&0.992&1.0\\ \hline
					{\bf Dishwasher}&0.021&0.0&0.0&0.062&0.0&0.117&0.353&0.554&0.699\\ \hline
					{\bf Drying machine}&0.003&0.0&0.067&0.0&0.0&0.03&0.14&0.287&0.495\\ \hline
					{\bf Electric stove}&0.961&1.0&1.0&1.0&1.0&0.983&0.994&0.997&1.0\\ \hline
					{\bf Electricity}&0.998&1.0&1.0&1.0&1.0&0.999&1.0&0.999&1.0\\ \hline
					{\bf Freezer}&0.828&1.0&0.733&0.375&1.0&0.835&0.913&0.979&1.0\\ \hline
					{\bf Fridge}&0.999&1.0&1.0&1.0&1.0&1.0&1.0&1.0&1.0\\ \hline
					{\bf Hair dryer}&0.895&0.889&1.0&0.938&1.0&0.977&0.994&1.0&1.0\\ \hline
					{\bf Internet}&0.132&1.0&1.0&1.0&1.0&0.668&0.911&0.974&1.0\\ \hline
					{\bf Iron}&0.977&1.0&1.0&0.938&1.0&0.998&0.999&1.0&1.0\\ \hline
					{\bf Microwave}&0.066&0.0&0.067&0.125&0.0&0.324&0.62&0.829&0.971\\ \hline
					{\bf Mobile phone}&0.817&1.0&1.0&1.0&1.0&0.979&0.998&1.0&1.0\\ \hline
					{\bf Motorcycle or scooter}&0.017&0.0&0.0&0.0&0.0&0.055&0.16&0.366&0.728\\ \hline
					{\bf Non-mobile phone}&0.877&1.0&0.867&0.562&0.889&0.918&0.965&0.99&0.99\\ \hline
					{\bf Owns dwelling}&0.879&0.444&0.8&0.312&1.0&0.788&0.894&0.967&0.99\\ \hline
					{\bf Owns land}&0.346&0.0&0.0&0.0&0.0&0.313&0.44&0.679&0.99\\ \hline
					{\bf Owns animals}&0.231&0.0&0.0&0.0&0.0&0.196&0.271&0.458&0.825\\ \hline
					{\bf PC/laptop}&0.188&1.0&1.0&1.0&1.0&0.738&0.957&0.997&1.0\\ \hline
					{\bf Radio}&0.814&0.0&0.0&0.0&0.0&0.812&0.891&0.932&0.971\\ \hline
					{\bf Table with chairs}&0.997&1.0&0.933&1.0&0.889&0.998&1.0&1.0&1.0\\ \hline
					{\bf Television}&0.995&1.0&1.0&1.0&1.0&0.999&0.999&0.999&1.0\\ \hline
					{\bf Tractor}&0.092&0.0&0.0&0.0&0.0&0.12&0.2&0.401&0.883\\ \hline
					{\bf Truck}&0.004&0.0&0.0&0.0&0.0&0.009&0.029&0.085&0.291\\ \hline
					{\bf Vacuum cleaner}&0.955&1.0&1.0&1.0&1.0&0.993&0.996&0.999&1.0\\ \hline
					{\bf Wardrobe}&0.998&1.0&1.0&1.0&1.0&0.998&1.0&1.0&1.0\\ \hline
					{\bf Washing machine}&0.947&1.0&0.933&0.938&1.0&0.992&0.996&0.997&1.0\\ \hline
					{\bf Watch}&0.575&0.111&0.933&0.062&1.0&0.715&0.849&0.928&0.971\\ \hline
					{\bf Water heater}&0.97&0.889&1.0&1.0&1.0&0.988&0.993&0.996&0.99\\ \hline
				\end{tabular}
		\end{center}
		\caption{Proportion of households in nodes (5,0)-(9,0) answering each of the 34 questions in the affirmative.}
		\label{table:itemspernode2}
	\end{figure}


\clearpage

\bibliographystyle{amsplain}



\end{document}